\documentclass[10pt,showpacs,letterpaper,aps,prl,twocolumn]{revtex4-1}
\usepackage[draft]{hyperref}
\usepackage{braket}
\usepackage{amsmath}
\usepackage{graphicx}
\usepackage{amssymb}
\usepackage{esint}
\usepackage{comment}

\newcommand{\hc}{\text{h.c.}}

\newcommand{\cc}{\text{c.c.}}
\newcommand{\vac}{\text{vac}}
\newcommand{\supp}{Supplementary Material }

\begin{document}

\title{Time ordering effects in the generation of \\
entangled photons using nonlinear optical processes}
\author{Nicol\'as Quesada and J. E. Sipe}
\affiliation{McLennan Physical Laboratories, University of Toronto, 60 St. George Street, Toronto, ON, M5S1A7, Canada }
\email{nquesada@physics.utoronto.ca}

\begin{abstract}
We study the effects of time ordering in photon generation processes such as
spontaneous parametric down-conversion (SPDC) and four wave mixing (SFWM). 
The results presented here are used to construct an intuitive picture that allows us to predict when time ordering
effects significantly modify the joint spectral amplitude (JSA) of the photons generated
in SPDC and SFWM. These effects become important only when the photons being 
generated lie with the pump beam that travels through the non-linear material for a significant amount of time. 
Thus sources of spectrally separable photons are ideal candidates for the observation of 
modifications of the JSA due to time ordering.
\end{abstract}

\maketitle

\preprint{APS/123-QED} 

\affiliation{McLennan Physical Laboratories, University of Toronto, 60 St.
George Street, Toronto, ON, M5S1A7, Canada }

\emph{Introduction} --- Particles exhibiting strong, non-classical
correlations play a central role in the implementation of quantum-enhanced
information processing tasks. \ Nonlinear photonic materials provide some of
the most versatile sources of these particles. \ The interactions that give
rise to the quantum correlated photons are typically rather weak, but by
using a strong classical pump field they can be enhanced by orders of
magnitude. \ Thus the theoretical treatment of photon generation usually
involves a time dependent Hamiltonian in the interaction picture. \ Because
this Hamiltonian generally does not commute with itself at different times,
the time evolution operator in these types of problems is not just a simple
exponential involving the time integral of the Hamiltonian, and so-called
``time ordering'' corrections arise in the description of the quantum state
evolution \cite{grice97,aggie10,aggie11}.

These time ordering effects can be ignored if the pump pulse is weak enough;
to first order in perturbation theory the interaction Hamiltonian appears
only once, and there is no need to worry about the time ordering of two
Hamiltonians \cite{grice97}. But as sources are improved and we move away
from this perturbative regime, time ordering effects start to become an
important issue that must be addressed. \ One strategy to deal with them is
to use the Dyson series \cite{aggie10,aggie11} for the evolution operator.
\ As discussed elsewhere \cite{nico14}, this method has the drawback that
the resulting approximate evolution is not unitary if the series is
truncated at any finite order. Perhaps more importantly, the method 
\emph{does not preserve the photon statistics} that the exact solution of the
problem is known to have, and thus leads to unphysical predictions. An
approach to achieving the correct photon statistics was proposed by Christ
et al.\cite{christ13n}, involving a \emph{numerical} method that seeks
solutions of the Heisenberg equations of motion satisfying the
\textquotedblleft right\textquotedblright\ photon statistics.

In this letter, we introduce an \emph{analytical} model that describes the
effects of time ordering in several non-linear quantum optical processes.
The model \emph{automatically captures the photon statistics and unitary
evolution that characterize the exact solution of the problem}. With this
model we provide a simple explanation of when time ordering effects become
important: \ As long as the generated photons do not significantly
copropagate (\emph{i.e.,} reside in the same spatial region at the same
time) with the pump photons, time ordering effects are irrelevant for the
description of the generated quantum state.

\emph{Model} --- We consider the following interaction picture Hamiltonian,
which can be used to model both spontaneous parametric down-conversion
(SPDC) and spontaneous four-wave mixing (SFWM) in a one dimensional
geometry, within the undepleted pump and rotating wave approximations: 
\begin{eqnarray}
\hat{H}_{I}(t)=-\hbar \varepsilon \int d\omega _{p}d\omega _{a}d\omega _{b}
&&e^{i\Delta t}\Phi (\omega _{a},\omega _{b},\omega _{p})  \label{hamil1} \\
&&\alpha (\omega _{p})\hat{a}^{\dagger }(\omega _{a})\hat{b}^{\dagger
}(\omega _{b})+\text{h.c.}  \notag
\end{eqnarray}%
Here $a^{\dagger }(\omega )$ and $b^{\dagger }(\omega )$ are creation
operators for two different modes (perhaps waveguide modes), and $\Phi $ is
the phase matching function, which we take to be a Gaussian function, 
\begin{equation}
\Phi (\omega _{a},\omega _{b},\omega _{p})=\exp \left( -(s_{a}\delta \omega
_{a}+s_{b}\delta \omega _{b}-s_{p}\delta \omega _{p})^{2}\right) .
\end{equation}%
When we take $s_{i}=\sqrt{\gamma }L/(2v_{i})>0$, where $v_{i}$ is the group
velocity at frequency $\omega _{i}$, this provides a good approximation to
the usual sinc function resulting from a uniform nonlinearity over a region
of length $L$, under the neglect of dispersion effects; the choice of $\gamma
=0.193$ guarantees that the full width at half maximum of the sinc and
Gaussian functions agree.  \ Here $\delta \omega _{i}=\omega _{i}-\bar{\omega%
}_{i}$; in SPDC the central frequencies $\bar{\omega}_{i}$ are constrained
by energy conservation, $\bar{\omega}_{a}+\bar{\omega}_{b}=\bar{\omega}_{p}$
, and momentum conservation$,$ $k_{a}(\bar{\omega}_{a})+k_{b}(\bar{\omega}%
_{b})=k_{p}(\bar{\omega}_{p})$, while for SFWM $\omega _{p}$ is \textit{twice} the frequency of the pump and momentum conservation reads $k_{a}(\bar{%
\omega}_{a})+k_{b}(\bar{\omega}_{b})=2k_{p}(\bar{\omega}_{p}/2)$ \cite{nico14}. \ The parameter $\varepsilon $ is a dimensionless constant that
characterizes the strength of the interaction\cite{nico14}; it depends on physical
properties such as the effective non-linearity of the medium and the area of
the interaction region. \ We also assume that the classical pump can be
described by a Gaussian function, 
\begin{equation}\label{pump}
\alpha (\omega _{p})=\frac{\tau }{\sqrt{\pi }}\exp \left( -\tau ^{2}\delta
\omega _{p}^{2}\right) ,
\end{equation}%
with $\tau $ characterizing the temporal duration of the pump pulse. Finally 
$\Delta =\omega _{a}+\omega _{b}-\omega _{p}$, and in Eq. (\ref{hamil1}) we
adopt the convention (used throughout the paper) that whenever the limits of
an integral are not specified they range from $-\infty $ to $\infty $.

Using the Magnus expansion, it was recently shown \cite{nico14} that the
unitary evolution operator generated by the Hamiltonian (\ref{hamil1}) can
be factored as $\hat{\mathcal{U}}=\hat{\mathcal{U}}_{\text{sq}}\hat{\mathcal{%
U}}_{\text{fc}}$ with $\hat{\mathcal{U}}_{\text{sq}}$ a pure two-mode
squeezing term (generated by terms that are extensions of $\hat{a}^{\dagger }%
\hat{b}^{\dagger }+\text{h.c.}$) and $\hat{\mathcal{U}}_{\text{fc}}$ a pure
single mode frequency conversion term (generated by terms that are
extensions of $a^{\dagger }a+b^{\dagger }b$), 
\begin{eqnarray}
\hat{\mathcal{U}} &=&e^{-2\pi i\int d\omega _{a}d\omega
_{b}\sum_{n}(J_{n}-iK_{n})\hat{a}^{\dagger }(\omega _{a})\hat{b}^{\dagger
}(\omega _{b})+\text{h.c.}}\times   \label{unitary} \\
&&e^{-2\pi i\sum_{c=\{a,b\}}\int d\omega _{c}d\omega _{c^{\prime
}} \sum_{m} G_{m}^{c}(\omega _{c},\omega _{c^{\prime }})\hat{c}^{\dagger }(\omega _{c})%
\hat{c}(\omega _{c}^{\prime })},  \notag
\end{eqnarray}%
with $n$ odd and $m$ even. Each of these indices labels the order of the
expansion in $\varepsilon $; \emph{e.g.}, $n=1$ is a term $\propto
\varepsilon $ and $m=2$ is a term $\propto \varepsilon ^{2}$. If a
spontaneous process is considered, where the unitary operator $\hat{\mathcal{%
U}}$ is applied to the vacuum, it is readily seen that the second
exponential in (\ref{unitary}) does not modify the initial state; $\hat{%
\mathcal{U}}_{\text{fc}}\ket{\vac}=\ket{\vac}$, and hence the output state $%
\ket{\psi_{\text{out}}}=\hat{\mathcal{U}}\ket{\vac}=\hat{\mathcal{U}}_{\text{%
sq}}\ket{\vac}$ is generated solely by the squeezing part $\hat{\mathcal{U}}%
_{\text{sq}}.$ With these observations it is easy to identify $%
J=\sum_{n}(J_{n}-iK_{n})$ as the Joint Spectral Amplitude (JSA) of the
photons generated in modes $a$ and $b$ in a spontaneous process. The first
term of the Magnus expansion for the Hamiltonian (\ref{hamil1}) is obtained
by integrating the Hamiltonian from $-\infty $ to $\infty $; $K_{1}=0$, and
the result for $J_{1}$ can be written in the following simple form:\ 
\begin{eqnarray}
J_{1}(\delta \omega _{a},\delta \omega _{b}) &=&-\frac{\varepsilon \tau }{%
\sqrt{\pi }}\exp \left( -\mathbf{u}\mathbf{N}\mathbf{u}^{T}\right) ,
\label{j1pdc} \\
\mathbf{N} &=&\left( 
\begin{array}{cc}
\mu _{a}^{2} & \mu ^{2} \\ 
\mu ^{2} & \mu _{b}^{2} 
\end{array}%
\right) ,\quad \mathbf{u}=(\delta \omega _{a},\delta \omega _{b}), \\
\eta _{a,b}=s_{p}-s_{a,b} &,&\ \mu ^{2}=\tau ^{2}+\eta _{a}\eta _{b},\ \mu
_{a,b}^{2}=\tau ^{2}+\eta _{a,b}^{2}.  \notag
\end{eqnarray}%
If time ordering effects were ignored, $K$ would vanish, and $J_{1}$ would
be the only contribution to $J$; $J_{1}$ would thus completely describe the
properties of the state, and could be identified with the joint spectral
amplitude of the generated photons. A measure of entanglement is the Schmidt
number \cite{law04}, which for the state characterized by $J_{1}$ is \cite{lvov07,mauerer07} 
\begin{equation*}
S=\mu _{a}\mu _{b}/\sqrt{\mu _{a}^{2}\mu _{b}^{2}-\mu ^{4}}.
\end{equation*}%
In the limit that the matrix $\mathbf{N}$ is diagonal, $\mu \rightarrow 0$
and the down converted photons are uncorrelated; here $S=1$. If in this
limit $\mathbf{N}$ were also proportional to the identity matrix, then the
photons would not only be uncorrelated but would share the same spectral
characteristics; for this to happen $\eta _{a}=-\eta _{b}$. In the opposite
limit, when 
\begin{equation}\label{ent}
\tau \ll |\eta _{a,b}|,
\end{equation}
$\mathbf{N}$ becomes rank deficient, the photons generated in the modes $a$
and $b$ are strongly correlated, and $S\rightarrow \infty $. \ A simple way
to understand the physics in the limit (\ref{ent}) is to rewrite the
inequality as 
\begin{equation*}
v_{p}\tau /|v_{a,b}-v_{p}|\ll L/v_{a,b}.
\end{equation*}%
That is, in this limit the time the down converted photons spend with the
pump photons is much less than the time they spend in the nonlinear
structure. The down converted photons once created thus escape the nonlinear
structure without spending appreciable time in the same region as the pump
photons, and if the pump intensity is high enough to generate a number of
pairs of photons then the stream of pairs is spread over a long time;
equivalently, their frequencies are tightly constrained. \ 

\emph{Time ordering effects in spontaneous photon generation} --- For the
simple Gaussian model considered in this letter, the real part of the first
non vanishing correction to the JSA due to time ordering is given by 
\begin{eqnarray}
J_{3}(\omega _{a},\omega _{b}) &=&W(\omega _{a},\omega _{b})-Z(\omega
_{a},\omega _{b})V(\omega _{a},\omega _{b}), \\
W(\omega _{a},\omega _{b}) &=&\frac{2\pi ^{3/2}\varepsilon ^{3}\tau ^{3}}{%
3R^{2}}\exp \left( -\mathbf{u}\mathbf{Q}\mathbf{u}^{T}/R^{4}\right) , \\
Z(\omega _{a},\omega _{b}) &=&4\sqrt{\pi }\tau ^{3}\varepsilon ^{3}\exp
\left( -\mathbf{u}\mathbf{N}\mathbf{u}^{T}/3\right) , \\
V(\omega _{a},\omega _{b}) &=&\int_{0}^{\infty }dp\int_{0}^{\infty }dq\exp
\left( -(p,q)\mathbf{M}(p,q)^{T}\right) \times   \notag \\
&&\cos \left( 4\tau \eta _{ab}(\delta \omega _{a}q+\delta \omega _{b}p)/%
\sqrt{3}\right) ,
\end{eqnarray}%
with 
\begin{eqnarray}
\mathbf{Q} &=&\left( 
\begin{array}{cc}
M^{4}\mu _{a}^{2} & \mu ^{6} \\ 
\mu ^{6} & M^{4}\mu _{b}^{2}%
\end{array}%
\right) ,\quad \mathbf{M}=\left( 
\begin{array}{cc}
2\mu _{a}^{2} & \mu ^{2} \\ 
\mu ^{2} & 2\mu _{b}^{2}  
\end{array}%
\right) , \\
R^{4} &=&4\mu _{a}^{2}\mu _{b}^{2}-\mu ^{4}=4\eta _{ab}^{2}\tau
^{2}+3\mu^4,  \notag \\
M^{4} &=&4\mu _{a}^{2}\mu _{b}^{2}-3\mu ^{4},\quad \eta _{ab}=-\eta
_{ba}=\eta _{a}-\eta _{b}.  \notag
\end{eqnarray}%
The imaginary part of the correction $K_{3}$ is given by 
\begin{eqnarray}
K_{3}(\omega _{a},\omega _{b}) &=&-\varepsilon ^{3}\pi ^{3/2}\frac{\tau ^{3}%
}{R^{2}}\exp (-\mathbf{u}\mathbf{Q}\mathbf{u}^{T}/R^{4})\times  \\
&&\left( \text{erfi}(y_{ab})+\text{erfi}(y_{ba})\right) ,  \notag \\
y_{ab} &=&\sqrt{\frac{2}{3}}\frac{\tau \eta _{ba}\left( 2\delta \omega
_{a}\mu _{a}^{2}-\delta \omega _{b}\mu ^{2}\right) }{R^{2}\mu _{a}}
\end{eqnarray}%
and $\text{erfi}(z)=\text{erf}(iz)/i$, with $\text{erf}$ being the standard
error function. Details of the calculation are given in the \supp. A useful feature of using Gaussian functions to model both
the phase-matching function and the pump pulse is that estimates of the
relative sizes of the corrections due to time-ordering in SPDC and SFWM can
be easily extracted; this also holds for the time-ordering corrections to
frequency conversion. \ We demonstrate this all below, introducing figures
of merit which characterize the size of the time-ordering corrections
relative to the uncorrected prediction that would follow from the first
order Magnus calculation. \ Some of the mathematical details are relegated
to the Supplementary Material.  

To estimate the effects of time ordering on SPDC and SFWM we introduce the
following figure of merit, 
\begin{equation*}
r=\frac{\max_{\omega _{a},\omega _{b}}|J_{3}(\omega _{a},\omega _{b})|}{%
\max_{\omega _{a},\omega _{b}}|J_{1}(\omega _{a},\omega _{b})|}=\frac{\sqrt{%
\pi }}{\varepsilon \tau }\max_{\omega _{a},\omega _{b}}|J_{3}(\omega
_{a},\omega _{b})|.
\end{equation*}
In the \supp we show that 
\begin{eqnarray}
|J_3(\omega_a,\omega_b)| \leq \frac{2 \pi^{3/2} \varepsilon^3 \tau^3}{R^2},
\end{eqnarray}
and so we identify the bound
\begin{equation}
r\leq 2 \pi ^{2}\varepsilon ^{2}\left( \frac{\tau ^{2}}{R^{2}}\right) .
\label{rresult}
\end{equation}
There are two interesting limits of the important ratio $\tau ^{2}/R^{2}$
that appears in (\ref{rresult}). \ We find 
\begin{eqnarray}
\tau  &\ll &|\eta _{a,b}|\;\;\Rightarrow \;\frac{\tau ^{2}}{R^{2}}\sim \frac{%
1}{\sqrt{3}}\frac{\tau }{|\eta _{a}|}\frac{\tau }{|\eta _{b}|}\lll 1,
\label{limits} \\
\mu  &=&0 \Rightarrow \;\frac{\tau ^{2}}{R^{2}}=\frac{\tau%
}{2 |\eta_{ab}|}.  \notag
\end{eqnarray}
From this we can see that the importance of the time ordering corrections
depends strongly on the entanglement of the photons in modes $a$ and $b$,
and we can also understand this scaling physically. \ As discussed in the
previous section, if the generated photons are highly entangled with a large
Schmidt number, the time they overlap the pump field is short. \ We would
then expect on physical grounds that time ordering corrections should be
small, and indeed\ we see from the first of (\ref{limits}) and (\ref{rresult}%
) that a very large interaction strength $\varepsilon $ would be necessary
for the time ordering corrections to come into play. In the reverse limit,
the second of (\ref{limits}), the group velocities of the generated photons satisfy $\eta_a \eta_b=-\tau^2$ ($\mu=0$). To have a significant time ordering correction, we have from the second of (\ref{limits}) that the group velocities of the downconverted photons must be similar. Note that if the velocities were equal $J_3$ would vanish identically; see the \supp. 
It can also be shown that the maximum of $\tau^2/R^2$ under the constraint of having $\mu=0$ is obtained for 
\begin{eqnarray}\label{sepid}
\eta_a=-\eta_b=\pm \tau \Rightarrow \frac{v_p \tau}{|v_{a,b}-v_p|}=\frac{\sqrt{\gamma}}{2}\frac{L}{v_{a,b}}
\end{eqnarray} 
then  $\tau^2/R^2$ equals $1/4$. Eq. (\ref{sepid}) simply tell us that the time the downconverted photons spend with the pump is comparable with the time they spend in the crystal.
Note that the entanglement in the downconverted state as predicted by using $J_{1}$ for $J$ and the condition (\ref{sepid}) is very small, but because of the continued presence of photons being generated with the pump pulse, time-ordering corrections would be physically expected to come into play at a smaller interaction strength $\varepsilon $; this is confirmed by using (\ref{sepid}) in (\ref{rresult}). Furthermore, besides being simply larger, time ordering corrections can be
expected to be of greater significance here, since they can qualitatively
modify the very small entanglement that the approximation of $J$ by $J_{1}$
would imply. \ 

When appropriate values of $\varepsilon $ are used in (\ref{rresult}), we
see that our results are consistent with what was found by Christ et al. \cite{christ13n}, in which simulations showed that extremely high pump
intensities are necessary to obtain observable time ordering effects.

\emph{Time ordering effects in stimulated photon generation} --- We now turn
to stimulated processes, in which the unitary operator $\mathcal{U}_{\text{fc%
}}$ in (\ref{unitary}) characterizing single mode frequency conversion
becomes important. The first contribution to the frequency conversion
unitary is given by 
\begin{eqnarray}
G_{2}^{a}(\omega _{a},\omega _{a}^{\prime }) &=&\varepsilon ^{2}\sqrt{\frac{%
\pi }{2}}\frac{\tau ^{2}}{\mu _{b}}e^{-x_{-}^{2}-x_{+}^{2}}\text{erfi}%
(x_{+}),  \label{g2pdc} \\
x_{+} &=&\frac{\tau \eta _{ba}\left( \delta \omega _{a}^{\prime }+\delta
\omega _{a}\right) }{\sqrt{2}\mu _{b}},\quad x_{-}=\frac{\left( \omega
_{a}-\omega _{a}^{\prime }\right) \mu _{a}}{\sqrt{2}}.  \notag
\end{eqnarray}%
An analogous formula can be found for $G_{2}^{b}(\omega _{b},\omega
_{b}^{\prime })$ by simply switching $a\leftrightarrow b$ in the above
formula. \ We consider the scenario of a process seeded by a coherent state
in mode $a$, described by the state 
\begin{equation*}
\ket{f_a}=\exp \left( i\int d\omega _{a}\left( f_{a}(\omega _{a})a^{\dagger
}(\omega _{a})+\text{h.c.}\right) \right) \ket{\vac}.
\end{equation*}%
Using (\ref{unitary}) we see that the lowest (second) order effect
associated with time ordering is \cite{nico14} 
\begin{eqnarray}
\ket{\psi_{\text{out}}} &=&\mathcal{U}_{\text{sq}}\mathcal{U}_{\text{fc}}%
\ket{f_a}=\mathcal{U}_{\text{sq}}\times  \label{secondorder} \\
&&e^{i\int d\omega _{a}(f_{a}(\omega _{a})+\delta f_{a}(\omega
_{a}))a^{\dagger }(\omega _{a})+\text{h.c.}}\ket{\vac},  \notag \\
\delta f_{a}(\omega _{a}) &=&-2\pi i\int d\omega _{a}^{\prime }f_{a}(\omega
_{a}^{\prime })G_{2}(\omega _{a},\omega _{a}^{\prime }).
\end{eqnarray}%
The last equation can be interpreted as follows: the first effect ($\propto
\varepsilon $) of the interaction is that the nonlinear medium acts as a two
mode squeezer, generating pairs of photons in modes $a$ and $b$. The second
order effect ($\propto \varepsilon ^{2}$), which appears solely due to time
ordering, describes the ``dressing'' of the seed photons by the pump. \ That
is, the frequency amplitudes of the photons of polarization $a$ are no
longer described by $f_{a}(\omega _{a})$ but rather by $f_{a}(\omega
_{a})+\delta f_{a}(\omega _{a})$. Finally, note that if instead the seed
were prepared in a single photon state $\ket{1(f_a)}=\int d\omega
_{a}f_{a}(\omega _{a})a^{\dagger }(\omega _{a})\ket{\vac}$, the effect of
time ordering would be to dress it according to $\ket{1(f_a)}\rightarrow %
\ket{1(f_a+\delta f_a)}$.

To quantify the effects of time ordering we estimate the correction given by
the second term of (\ref{secondorder}) relative to the first term. If we
take $f_{a}(\omega _{a})=\nu \exp (-\tau _{a}^{2}\delta \omega _{a}^{2})$,
in the limit $\tau _{a}\ll \tau ,|\eta _{a}|$, 
\begin{eqnarray}
\int& d\omega _{a}^{\prime }f_{a}(\omega _{a}^{\prime })G_{2}(\omega
_{a},\omega _{a}^{\prime }) =\varepsilon ^{2}\nu \frac{\pi ^{2}\tau ^{2}}{%
\mathcal{N}^{2}}\exp (-q^{2})\text{erfi}(q),  \notag \\
 &q=\sqrt{2}\omega _{a}\eta _{ab}\tau \mu _{a}/\mathcal{N}^{2},\quad 
\mathcal{N}^{4}=\mu ^{4}+2\tau ^{2}\eta _{ab}^{2}.
\end{eqnarray}%
Note that the maximum value the function $\exp (-q^{2})\text{erfi}(q)$
attains is $\xi \approx 0.610503$ and that the function is identically zero
if $q=0$ (which only occurs for the unphysical situation $\eta _{a}=\eta _{b}
$). \ As before, we can compare the relative strengths of the amplitudes of
the coherent state with and without including time ordering effects to
obtain the following figure of merit: 
\begin{equation}\label{rhoest}
\rho =\frac{\max_{\omega _{a}}|\delta f_{a}(\omega _{a})|}{\max_{\omega
_{a}}f_{a}(\omega _{a})}=2\pi ^{3}\xi (\varepsilon ^{2}\tau ^{2}/\mathcal{N}%
^{2}).
\end{equation}%
This quantity measures how much the pump is dressed by the time-ordering
effects of the stimulated process. It has a very simple interpretation: \ To
have important time ordering effects the seed photons, and the $a$ and $b$
photons created by the nonlinear process, must spend a significant amount of
time traveling with the pump pulse. \ Note in particular that if the photons
of type $a$ move significantly faster than those of type $b$, for example,
then $|\eta _{ab}|$ would make the bound (\ref{rhoest}) go to zero; if on
the other hand the $a$ photons travel much faster than those of the pump,
then $\eta _{a}\gg \tau \text{ and }\eta _{b}\sim \tau \Rightarrow \mu
^{2}=\tau ^{2}+\eta _{a}\eta _{b}\gg \tau ^{2}$ and then again $\rho
\rightarrow 0$.

\emph{Time ordering effects in photon conversion} --- For frequency
conversion (FC), a strong pump pulse is used to mediate the conversion of
photons from, for example, a lower frequency to photons of a higher
frequency. The Hamiltonian governing such processes for a one dimensional
geometry is 
\begin{eqnarray}
\hat{H}_{I}(t)=-\hbar \varepsilon \int d\omega _{p}d\omega _{a}d\omega _{b}
&&e^{i\tilde{\Delta}t}\tilde{\Phi}(\omega _{a},\omega _{b},\omega _{p})
\label{hamilFC} \\
&&\tilde{\alpha}(\omega _{p})\hat{a}(\omega _{a})\hat{b}^{\dagger }(\omega
_{b})+\text{h.c.},  \notag
\end{eqnarray}%
with $\tilde{\Delta}=\omega _{b}-\omega _{a}-\omega _{p}$. As in the
previous section, we consider Gaussian approximations for the phase matching
function 
\begin{equation}
\tilde{\Phi}(\omega _{a},\omega _{b},\omega _{p})=\exp \left( -(s_{b}\delta
\omega _{b}-s_{p}\delta \omega _{p}-s_{a}\delta \omega _{a})^{2}\right) ,
\end{equation}%
and the pump function as given in equation (\ref{pump}). Note that the
Hamiltonian (\ref{hamilFC}) can be obtained from equation (\ref{hamil1}) by
changing $a(\omega _{a})\Leftrightarrow a^{\dagger }(\omega _{a})$ and
switching the sign of all the scalars associated with $a$. Using the Magnus
expansion it can be shown that the time evolution operator $\hat{\mathcal{U}}
$ for frequency conversion can be factorized in terms of a two mode
frequency conversion term $\hat{\mathcal{U}}_{\text{2fc}}$ (generated by
terms that are extensions of $\hat{a}\hat{b}^{\dagger }+\text{h.c.}$)
preceded by a pure single mode frequency conversion term $\hat{\mathcal{U}}_{%
\text{fc}}$ (generated by terms that are extensions of $a^{\dagger
}a+b^{\dagger }b$), 
\begin{eqnarray}\label{uufc}
\hat{\mathcal{U}} &=&e^{-2\pi i\int d\omega _{a}d\omega _{b}\sum_{n}\tilde{J}%
_{n}(\omega _{a},\omega _{b})\hat{a}(\omega _{a})\hat{b}^{\dagger }(\omega
_{b})+\text{h.c.}}\times   \notag  \label{uufc} \\
&&e^{-2\pi i\sum_{c\in \{a,b\}}\int d\omega _{c}d\omega _{c}^{\prime }\tilde{%
\sum_{m}}\tilde{G}_{m}^{c}(\omega _{c},\omega _{c}^{\prime })\hat{c}%
^{\dagger }(\omega _{c})\hat{c}(\omega _{c}^{\prime })}.
\end{eqnarray}%
For the Gaussian functions assumed here for the phase-matching and the pump
pulse, we have $\tilde{J}_{1}(\delta \omega _{a},\delta \omega
_{b})=J_{1}(-\delta \omega _{a},\delta \omega _{b})$ and $\tilde{G}%
_{2}^{c}(\omega _{c},\omega _{c}^{\prime })=G_{2}^{c}(\omega _{c},\omega
_{c}^{\prime })$. \ Were the sum in the first exponential in (\ref{uufc})truncated at the
first term, and the second exponential ignored, we would obtain the sum of
the Taylor series associated with the Hamiltonian (\ref{hamilFC}). If the
Hamiltonian commuted with itself at different times, this would be the exact
solution of the problem.\ Assuming time ordering effects are irrelevant, the
unitary $\hat{\mathcal{U}}_{\text{2fc}}$ can be written, via the Schmidt
decomposition, as a multimode beam-splitter that connects modes $\hat{A}_{n}$
and $\hat{B}_{n}$ 
\begin{eqnarray}
\hat{\mathcal{U}}_{\text{2fc}} &=&e^{-i\sum_{j=0}^{\infty }g_{j}(\hat{A}_{j}%
\hat{B}_{j}^{\dagger }+\text{h.c.})}=\prod_{n=0}^{\infty }e^{-ig_{j}(\hat{A}%
_{j}\hat{B}_{j}^{\dagger }+\text{h.c.})},  \notag  \label{g_j} \\
\hat{A}_{j} &=&\int d\omega _{a}\frac{H_{j}(-\mu _{a}\delta \omega _{a}/%
\sqrt{S/2})e^{-\mu _{a}^{2}\delta \omega _{a}^{2}/S}}{\sqrt{2^{j}j!\sqrt{\pi
S/2}/\mu _{a}}}\hat{a}(\omega _{a}),  \notag \\
\hat{B}_{j} &=&\int d\omega _{b}\frac{H_{j}(\mu _{b}\delta \omega _{b}/\sqrt{%
S/2})e^{-\mu _{b}^{2}\delta \omega _{b}^{2}/S}}{\sqrt{2^{j}j!\sqrt{\pi S/2}%
/\mu _{b}}}\hat{b}(\omega _{b}),  \notag \\
g_{j} &=&\theta _{0}\sqrt{1+s^{2}}s^{j},\quad \theta _{0}=2\pi \varepsilon
\tau /\sqrt{2\mu _{a}\mu _{b}}.
\end{eqnarray}%
where the $H_{j}(x)$ are Hermite polynomials of degree $j$, the quantity $s$
is implicitly defined via $S=(1+s^{2})/(1-s^{2})$, and $0\leq s\leq 1$; note
that the operators $A_{j},B_{j}$ satisfy the usual bosonic commutation
relations $[A_{j},A_{k}^{\dagger }]=[B_{j},B_{k}^{\dagger }]=\delta _{jk}$,
with all other commutators vanishing. To convert a coherent state (or a
single photon) of type $a$ and frequency centered at $\bar{\omega}_{a}$ to
one of type $b$ centered at frequency $\bar{\omega}_{b}$ one chooses an $n$
-- usually $n=0$ is the simplest choice, since it corresponds to a unimodal
Gaussian profile -- \ then prepares the state $\ket{A_n}=e^{i\alpha (\hat{A}%
_{n}+\hat{A}_{n}^{\dagger })}\ket{\vac}(\ket{1_{A_n}}=\hat{A}_{n}^{\dagger }%
\ket{\vac})$, and tunes the pump and nonlinear medium in such a way that $%
g_{n}=\pi /2$. This will guarantee that the output state is $\ket{B_n}%
=e^{i\alpha (\hat{B}_{n}+\hat{B}_{n}^{\dagger })}\ket{\vac}=\hat{\mathcal{U}}%
_{\text{2fc}}|_{g_{n}=\pi /2}\ket{A_n}$ ($\ket{1_{B_n}}=B_{n}^{\dagger }%
\ket{\vac}=\hat{\mathcal{U}}_{\text{2fc}}|_{g_{n}=\pi /2}\ket{1_{A_n}}$).

As it is clear from Eq. (\ref{uufc}), time ordering effects will lead to two
modifications of this scenario. The first is that the conversion amplitude $%
\tilde{J}(\omega _{a},\omega _{b})$ is no longer $\tilde{J}_{1}(\omega
_{a},\omega _{b})$, but involves corrections $\tilde{J}_{n}(\omega
_{a},\omega _{b})$. The second and more important is that the unitary
connecting input and output is given by (\ref{uufc}), which contains not
only a two mode frequency conversion unitary \emph{but also} a single
frequency conversion term $\hat{\mathcal{U}}_{\text{fc}}$. \ To lowest
order, this correction is encoded in the functions $\tilde{G}_{2}^{c}(\omega
_{c},\omega _{c}^{\prime })=G_{2}^{c}(\omega _{c},\omega _{c}^{\prime })$ in
(\ref{uufc}). Because of this equality, the action of the second order
Magnus correction on a coherent or single photon state is identical to (\ref%
{secondorder}), and we can obtain a bound on the time ordering effects which
is identical to (\ref{rhoest}) and has the same interpretation: \ Time
ordering effects are unimportant if the pump and frequency converted photons
do not significantly copropagate.

\begin{table}[!t]
\begin{tabular}{l|l|l|l|l|l}
Mode & Central & Central & Index of & Group & Polari- \\ 
label & frequency & wavelength & Refraction & velocity & zation \\ 
\hline\hline
$p$ & 2.0000 PHz & 0.9418 $\mu$m & 2.15541 & $c$/2.20054 & e \\ 
$a$ & 1.2707 PHz & 1.4824 $\mu$m & 2.21112 & $c$/2.26276 & o \\ 
$b$ & 0.7293 PHz & 2.5827 $\mu$m & 2.10269 & $c$/2.17833 & e \\ \hline
\end{tabular}
\caption{Parameters for PPLN to observe significant time ordering effects.
To obtain zeroth order phase-matching a periodicity of $\Lambda=1/|n_e(%
\protect\lambda_p)/\protect\lambda_p-n_o(\protect\lambda_a)/\protect\lambda%
_a-n_e(\protect\lambda_b)/\protect\lambda_b|=58.25 \protect\mu$m is
required. $n_{o/e}$ are the indices of refraction of the two different
polarizations.}
\label{table1}
\end{table}

\begin{figure*}[t]
\caption{First and third order Magnus terms for the parameters from Table I. Note that the function $J_1$ as been normalized by $\varepsilon \tau$ and $J_3, K_3$ by $\varepsilon^3 \tau$.}
\label{fig1}\includegraphics[width=\textwidth]{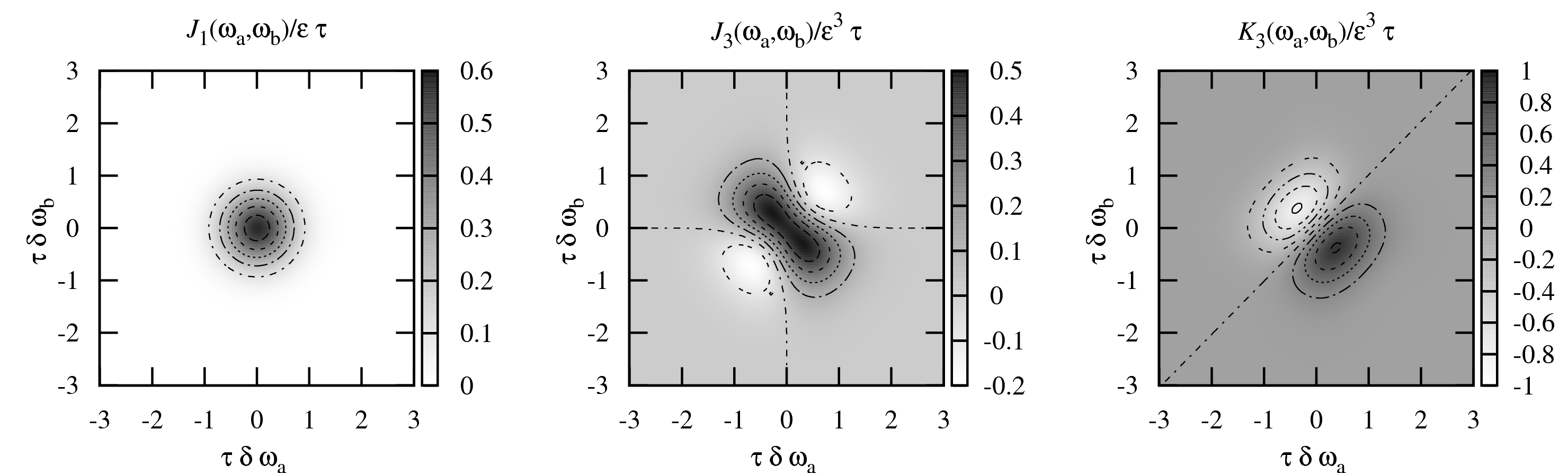} 
\end{figure*}

\emph{Experimental proposal} --- Although in many applications \cite%
{resch1,Walmsley11} a simple perturbative treatment of the nonlinear
processes described here is appropriate, there are at least two experiments%
\cite{lang05,van04} in which this approximation breaks down and a non
perturbative treatment is needed. In both experiments a periodically poled
Lithium Niobate (PPLN) crystal was used to upconvert a photon from the telecommunication wavelength to the visible with near unit efficiency.
The fact that the single photon is converted with nearly unit efficiency
implies that one of the $g_{j}$ in (\ref{g_j}) is roughly equal to $\pi /2$;
that is, the nonlinear medium acts as a frequency beam splitter with
reflectivity approaching unity. This immediately implies that $\varepsilon
\sim 1$. \ One could question whether or not in those experiments time
ordering effects are important. It turns out that they are not, simply
because non-degenerate Type-0 phase matching was used. Since there is only
one polarization available and the group velocity curve as a function of
frequency is a monotonic function, it is impossible to have pump and down
converted photons travelling together for a significant amount of time, and $%
\tau ^{2}/R^{2}\ll 1$. \  Nevertheless, it is possible to make PPLN
birefringent by adding small amounts of Magnesium Oxide (MgO). In the study
of Gayer et al.\cite{gayer08}, Sellmeier equations were determined for the
ordinary and extraordinary polarizations of PPLN doped with MgO in the
wavelength window 0.5 to 4 nm. Although in the experiments mentioned so
far a PPLN crystal was used for FC, they could also be used for SPDC. In
Table \ref{table1} we show parameters for which it is possible to have pump
and down converted photons travelling together. Using the parameters
described in Table \ref{table1} it would be possible to set up an experiment
in which, for the first time, a JSA that is nonlinear in the pump electric
field could be observed. Using the parameteters from Table \ref{table1},
with $\chi _{2}=10$ pm/V, a crystal length of $L=4$ cm, a pulse duration 
$\tau =1$ ps, and a peak electric field of $6$ MV/m, we obtain 
\begin{equation*}
\varepsilon =0.30\text{ and }r\leq 2\pi ^{2}\varepsilon ^{2}\frac{\tau ^{2}%
}{R^{2}}=0.46.
\end{equation*}%
As it is seen in Fig. \ref{fig1} time ordering effects (encoded in $J_{3}$
and $K_{3}$) are comparable to the first order Magnus term $J_{1}$. These
modifications could open an avenue for the generation of squeezed states
with very interesting JSAs, and in any case these effects will become
important as soon as very squeezed states are generated.

\emph{Conclusion} --- In this letter we have presented a theory that allows
us to understand the effects of time ordering in nonlinear quantum optics
processes. \ In the last three sections we investigated the effects of time
ordering for spontaneous and stimulated photon generation, and frequency
conversion. In each case a figure of merit quantifying the effects of time
ordering was introduced, and it was shown that these effects are only
sizable whenever the photons in the down-converted/frequency converted modes
copropagate with the pump photons. This agrees with the simple physical
intuition that the only way the effect of the Hamiltonian at some posterior
time $\hat{H}(t_{f})$ can be influenced by the former version of itself $%
\hat{H}(t_{i})$, under the assumption of an undepleted pump, is if the
fields associated with the pump photons, any seed photons, and the generated
photons spend a significant amount of time in the same regions of space. \
Under these conditions the time ordering corrections will also have the most
significant qualitative effect, since they will modify the Schmidt number of
the quantum correlated photons from the uncorrected value of close to unity.
\ 

We have used Gaussian functions to approximately describe both the pump
pulse and the phase-matching functions that enter the description of the
photon generation or conversion. \ This led to the possibility of
constructing analytic expressions for the figures of merit. \ Nonetheless,
the general approach of the Magnus expansion described earlier \cite{nico14}
could be applied to the use of arbitrary functions to model the pump pulse
and phase matching. \ A general result of our investigations here, that
strong pump pulses are required for any deviation from the first order
Magnus term to be relevant, can be expected to hold more generally. \ There
is thus a wide range of problems where an elementary first order
perturbative approach would fail, but a calculation using the first order
Magnus term would be sufficient. \ Yet working at pump intensities high
enough that time ordering corrections appear should lead to new strategies
for producing novel quantum states. \  

\appendix

\begin{widetext}
\section{\supp for ``Time ordering effects in the generation of entangled photons using nonlinear optical processes''}
\noindent In this \supp we calculate the third order Magnus term $\Omega_3$ \cite{nico14} and associated JSA contribution $J_3$ for the Hamiltonian considered in the main text. Note that the second order correction $\Omega_2$ (and associated functions $G_2^a, G_2^b$) and the imaginary part of the third order correction $K_3$ can be calculated directly from the results in \cite{nico14}. For the calculation of $J_3$ we will not use the results presented in \cite{nico14}. The reason is that integrals of the form $\fint \frac{dx}{x} \frac{dy}{y} f(x,y)$, which are the ones necessary to calculate the third order correction, are hard to bound and cannot be computed analytically for the situation when $f$ is a Gaussian function. To circumvent this difficulty we assume that the Hamiltonian can be written as a function of the frequencies of the down-converted photons and time ($\omega_a,\omega_b,t$) but without any explicit dependence of the pump frequency ($\omega_p$) which will be assumed to have been integrated out.  This can be done analytically for Gaussian pump profiles and phase-matching functions, as we shall show shortly. Because the time dependence of the Hamiltonian will no longer be harmonic but Gaussian, the time order integrations will not involve any principal value integrals but rather incomplete (\emph{i.e.} not extending over the whole real line) Gaussian integrals.

We remind the reader that the Hamiltonian considered in the main text is given by Eq. (1) and that the phase matching and pump profile functions are defined by Eq. (2) and (3) respectively. We also note that the factor $\Delta$ appearing in Eq. (1) can be written as
\begin{eqnarray*}
\Delta=\omega_a+\omega_b-\omega_p=\delta\omega_a+\delta\omega_b-\delta\omega_p. 
\end{eqnarray*}
Given the simplicity of the pump and phase matching functions we can perform analytically the integral over $\omega_p$
\begin{eqnarray}
F(\omega_a,\omega_b,t)&=&(-\varepsilon)\int d \omega_p e^{i \Delta t} \alpha(\omega_p) \Phi(\omega_a, \omega_b, \omega_p)=\\
&&(-\varepsilon) \tau  \sqrt{\frac{1}{s_p^2+\tau^2}} \exp\left(\frac{\left(-2 s_p \left(s_a \delta \omega_a+s_b \delta \omega _b\right)+i t\right){}^2}{4\left(s_p^2+\tau ^2\right)}+i t \left(\delta \omega _a+\delta \omega _b\right) -\left(s_a \delta \omega _a+s_b \delta \omega _b\right){}^2 \right).
\end{eqnarray}
After the integration the Hamiltonian is now
\begin{eqnarray}
H_I(t)=\hbar  \int d \omega_a d \omega_b F(\omega_a, \omega_b, t) a^\dagger (\omega_a) b^\dagger (\omega_b)+\hc
\end{eqnarray}
For the third order Magnus term we get
\begin{eqnarray}
\Omega_3&=&\frac{(-i)^3}{6 \hbar^3}\int dt \int^t dt' \int^{t'} dt''\left\{ \left[H_I(t),\left[H_I(t'),H_I(t'') \right] \right]+\left[\left[H_I(t),H_I(t') \right],H_I(t'') \right] \right\}\\
&=&\frac{i}{6}\int dt \int^t dt' \int^{t'} dt'' \int d \omega_o d \omega_e d \omega_o' d \omega_e' d \omega_o'' d \omega_e''   \nonumber \\
&& \times \Big( \bar \delta \left(\omega _e''-\omega _e\right) \bar \delta \left(\omega _o''-\omega _o'\right) F\left(\omega _o,\omega _e,t\right) F\left(\omega _o',\omega _e',t'\right) F\left(\omega _o'',\omega _e'',t''\right)^*  a^{\dagger }(\omega_o) b^\dagger(\omega _e')\nonumber \\
&&-2  \bar \delta \left(\omega _e'-\omega _e\right) \bar \delta \left(\omega _o'-\omega _o''\right) F\left(\omega _o,\omega _e,t\right) F\left(\omega _o'',\omega _e'',t''\right) F\left(\omega _o',\omega _e',t'\right)^* a^\dagger(\omega _o) b^\dagger(\omega_e'')\nonumber \\
&&+ \bar \delta \left(\omega _e''-\omega _e'\right) \bar \delta \left(\omega _o''-\omega _o\right) F\left(\omega _o,\omega _e,t\right) F\left(\omega _o',\omega _e',t'\right) F\left(\omega _o'',\omega _e'',t''\right)^* a^\dagger(\omega _o') b^\dagger(\omega _e)\nonumber \\
&&+ \bar \delta \left(\omega _e-\omega _e'\right) \bar \delta \left(\omega _o-\omega _o''\right) F\left(\omega _o',\omega _e',t'\right) F\left(\omega _o'',\omega _e'',t''\right) F\left(\omega _o,\omega _e,t\right)^*  a^\dagger(\omega_o') b^\dagger (\omega _e'')\nonumber \\
&&-2  \bar \delta \left(\omega _e'-\omega _e''\right) \bar \delta \left(\omega _o'-\omega _o\right) F\left(\omega _o,\omega _e,t\right) F\left(\omega _o'',\omega
   _e'',t''\right) F\left(\omega _o',\omega _e',t'\right)^* a(\omega _o'')^\dagger b^\dagger(\omega _e) \nonumber \\
&&+ \bar \delta \left(\omega _e-\omega _e''\right) \bar \delta \left(\omega _o-\omega _o'\right) F\left(\omega _o',\omega _e',t'\right) F\left(\omega _o'',\omega _e'',t''\right) F\left(\omega _o,\omega _e,t\right)^* a^\dagger(\omega _o'') b^\dagger(\omega _e')\nonumber \Big) -\hc 
\end{eqnarray}
In the last equation we use $\bar \delta$ for the Dirac distribution. To simplify this expression we perform the following changes of variables in each of the six terms in the last equation
\begin{eqnarray}
\begin{array}{cccccc}
\omega _e\to \omega _d & \omega _e''\to \omega _d & \omega _o'\to \omega _c & \omega _o''\to \omega _c & \omega _e'\to \omega _b & \omega _o\to \omega _a\\
\omega _e\to  \omega _d & \omega _e'\to \omega _d & \omega _o'\to \omega _c & \omega _o''\to \omega _c & \omega _e''\to \omega _b & \omega _o\to \omega _a\\
\omega _e'\to \omega _d & \omega_e''\to \omega _d & \omega _o\to \omega _c & \omega _o''\to \omega _c & \omega _e\to \omega _b & \omega _o'\to \omega _a\\
\omega _e\to \omega _d & \omega _e'\to \omega _d & \omega _o\to \omega _c & \omega _o''\to \omega _c & \omega _e''\to \omega _b & \omega _o'\to \omega _a\\
\omega _e'\to \omega _d & \omega _e''\to \omega _d & \omega _o\to \omega _c & \omega _o'\to \omega _c & \omega _e\to \omega _b & \omega _o''\to \omega _a\\
\omega _e\to \omega _d & \omega _e''\to \omega _d & \omega _o\to \omega _c & \omega _o'\to \omega _c & \omega _e'\to \omega _b & \omega _o''\to \omega _a,
\end{array}
\end{eqnarray}
to get
\begin{eqnarray}
\Omega_3&=&\frac{i}{6}\int dt \int^t dt' \int^{t'} dt'' \int d \omega_a d \omega_b d \omega_c d \omega_d  a^\dagger(\omega _a) b^\dagger(\omega _b) \times\\
&&\Big(F\left(\omega _a,\omega _d,t\right) F\left(\omega _c,\omega _b,t'\right) F\left(\omega _c,\omega _d,t''\right)^*-2 F\left(\omega _a,\omega _d,t\right) F\left(\omega _c,\omega _b,t''\right) F\left(\omega _c,\omega _d,t'\right)^*\nonumber \\ 
&& +F\left(\omega _a,\omega _d,t'\right) F\left(\omega _c,\omega _b,t\right) F\left(\omega _c,\omega _d,t''\right)^*+F\left(\omega _a,\omega _d,t'\right) F\left(\omega _c,\omega _b,t''\right) F\left(\omega _c,\omega_d,t\right)^*\nonumber \\ 
&& -2 F\left(\omega _a,\omega _d,t''\right) F\left(\omega _c,\omega _b,t\right) F\left(\omega _c,\omega _d,t'\right)^*+F\left(\omega _a,\omega _d,t''\right) F\left(\omega _c,\omega _b,t'\right) F\left(\omega _c,\omega _d,t\right)^*\Big)-\hc \nonumber.
\end{eqnarray}
We can now perform the change of variables $t=q+2 r+s, \ t'=q-r+s, \ t''=q-r-2 s $ that transforms the integral according to $\int dt \int ^{t} d t' \int ^{t'} dt'' = 9 \int dq \int_0^{\infty} dr \int_0^{\infty} ds$, and also
use the fact that $F(\omega_a, \omega_b, -t)=F(\omega_a,\omega_b,t)^*$, where $x^*$ denotes the complex conjugate of $x$, to obtain
\begin{eqnarray}
\Omega_3&=&\frac{3i}{2} \int dq \int_0^{\infty} dr \int_0^{\infty} ds \int d \omega_a d \omega_b d \omega_c d \omega_d  a^\dagger(\omega _a) b^\dagger(\omega _b) \times \\
&& \big(
F\left(\omega _a,\omega _d,q+2 r+s\right) F\left(\omega _c,\omega _b,q-r+s\right) F\left(\omega _c,\omega _d,-q+r+2 s\right)\nonumber \\
&&-2 F\left(\omega _a,\omega_d,q+2 r+s\right) F\left(\omega _c,\omega _b,q-r-2 s\right) F\left(\omega _c,\omega _d,-q+r-s\right)\nonumber \\
&&+F\left(\omega _a,\omega   _d,q-r+s\right) F\left(\omega _c,\omega _b,q+2 r+s\right) F\left(\omega _c,\omega _d,-q+r+2 s\right)\nonumber \\
&&+F\left(\omega _a,\omega _d,q-r+s\right) F\left(\omega _c,\omega _b,q-r-2 s\right) F\left(\omega _c,\omega _d,-q-2 r-s\right)\nonumber \\
&&-2 F\left(\omega _a,\omega _d,q-r-2 s\right) F\left(\omega _c,\omega _b,q+2 r+s\right) F\left(\omega _c,\omega _d,-q+r-s\right)\nonumber \\
&&+F\left(\omega _a,\omega _d,q-r-2 s\right) F\left(\omega _c,\omega _b,q-r+s\right) F\left(\omega _c,\omega _d,-q-2 r-s\right) \big) -\hc \nonumber.
\end{eqnarray}
Note the following: First, $r$ and $s$ can always be swapped since they have the same integration ranges and second the change  $q \to -q$ can always be performed since $\int_{-\infty}^{\infty} d q f(q)=\int_{-\infty}^{\infty} d q f(-q)$. Because of this it is easily seen that in the last equation the sixth term is the complex conjugate of the first, the fifth is the complex conjugate of the second and the fourth is the complex conjugate of the third. So we can more compactly write
\begin{eqnarray}
\Omega_3&=& -2 \pi i \int d \omega_a d \omega_b  a^\dagger(\omega _a) b^\dagger(\omega _b) J_3(\omega_a,\omega_b) -\hc\\
&=& -2 \pi i\int d \omega_a d \omega_b a^\dagger(\omega _a) b^\dagger(\omega _b)  \left( -\frac{3}{4 \pi} \int dq d \omega_c d \omega_d (L_1-2 L_2+L_3 +\cc) \right)-\hc\\
L_1&=& \int_0^{\infty} dr \int_0^{\infty} ds \ F\left(\omega _a,\omega _d,q+2 r+s\right) F\left(\omega _c,\omega _b,q-r+s\right) F\left(\omega _c,\omega _d,-q+r+2 s\right)\\
L_2&=& \int_0^{\infty} dr \int_0^{\infty} ds \ F\left(\omega _a,\omega_d,q+2 r+s\right) F\left(\omega _c,\omega _b,q-r-2 s\right) F\left(\omega _c,\omega _d,-q+r-s\right)\\
L_3&=& \int_0^{\infty} dr \int_0^{\infty} ds \ F\left(\omega _a,\omega   _d,q-r+s\right) F\left(\omega _c,\omega _b,q+2 r+s\right) F\left(\omega _c,\omega _d,-q+r+2 s\right).
\end{eqnarray}
To get a simpler expression we can perform the following changes of variables:
For $L_1$ in the last equation we put $r \to r'+s', s\to -s'$ to obtain:
\begin{eqnarray}
L_1=\int_{-\infty}^0 ds' \int_{-s'}^\infty dr' F\left(\omega _a,\omega_d,q+2 r'+s'\right) F\left(\omega _c,\omega _b,q-r'-2 s'\right) F\left(\omega _c,\omega _d,-q+r'-s'\right).
\end{eqnarray}
Likewise for $L_3$ we change $r \to -r'-s'$ and $s \to r'$ to obtain:
\begin{eqnarray}
L_3=\int_0^{\infty} dr' \int_{-\infty}^{-r'} ds' F\left(\omega _a,\omega_d,q+2 r'+s'\right) F\left(\omega _c,\omega _b,q-r'-2 s'\right) F\left(\omega _c,\omega _d,-q+r'-s'\right).
\end{eqnarray}
Note that now the integrands of $L_1, L_2$ and $L_3$ are identical and that $\int_{-\infty}^0 ds' \int_{-s'}^{\infty} dr'+\int_0^{\infty}dr'\int_{-\infty}^{-r'}ds'=\int_{0}^{\infty} dr' \int_{-\infty}^0 ds'$ so we can write more simply
\begin{eqnarray}
L_1-2L_2+L_3&=&\left(\int_{0}^{\infty} dr \int_{-\infty}^0 ds  -2\int_{0}^{\infty} dr \int_0^\infty ds \right) \times\\
&& F\left(\omega _a,\omega_d,q+2 r+s\right) F\left(\omega _c,\omega_b,q-r-2 s\right) F\left(\omega _c,\omega_d,-q+r-s\right) \nonumber,
\end{eqnarray}
we can now write $J_3$ as
\begin{eqnarray}
J_3(\omega_a,\omega_b)&=& -\frac{3}{4 \pi} \int dq d \omega_c d \omega_d \left(\int_{0}^{\infty} dr' \int_{-\infty}^0 ds  -2\int_{0}^{\infty} dr' \int_0^\infty ds \right) \times\\
&& F\left(\omega _a,\omega_d,q+2 r+s\right) F\left(\omega _c,\omega_b,q-r-2 s\right) F\left(\omega _c,\omega_d,-q+r-s\right) \nonumber.
\end{eqnarray}
Up to this point we have not used the fact that $F$ is Gaussian; now we will use it: The product of three $F$'s is a Gaussian function in $q,r,s,\omega_a,\omega_b,\omega_c,\omega_d$ and hence the integrals over $q, \omega_c, \omega_d$ are complete (\emph{i.e.} extending over the whole real line) Gaussian integrals and thus can be done analytically, leaving us with a Gaussian function in $r,s,\omega_a,\omega_b$. Doing these integrals we obtain
\begin{eqnarray}\label{bigeq}
\int dq d \omega_c d \omega_d   F\left(\omega _a,\omega_d,q+2 r+s\right) F\left(\omega _c,\omega_b,q-r-2 s\right) F\left(\omega _c,\omega_d,-q+r-s\right)= -\frac{2 \pi ^{3/2} \tau  \varepsilon^3 }{3 s_a s_b} e^{- \mathbf{u} \mathbf{N}\mathbf{u}^T/3  }  (f(r,s)+\cc), \nonumber
\end{eqnarray}
with
\begin{eqnarray}
f(r,s)=\exp(\mathbf{x}^T \mathbf{A} \mathbf{x}+i \mathbf{v}^T \mathbf{x}), \quad
\mathbf{A}=-\frac{3}{4 \tau^2}\left(\begin{array}{cc}  
 \frac{2\mu_b^2}{ s_b^2} & \frac{\mu^2}{ s_a s_b} \\
 \frac{\mu^2}{ s_a s_b} &  \frac{2\mu_a^2}{ s_a^2}  
\end{array}\right), \quad \mathbf{v}^T= 2 \eta_{ab }\left(\frac{ \delta \omega _a}{ s_b},\frac{  \delta \omega _b }{s_a}\right).
\end{eqnarray}
and $\mathbf{N}$ and $\mathbf{u}$ being defined in Eq. (6) of the main text.
 With this result we can write
\begin{eqnarray}
J_3(\omega_a,\omega_b)&=&  \frac{ \sqrt{\pi}  \tau  \varepsilon^3 }{2  s_a s_b} \exp \left(- \mathbf{u} \mathbf{N}\mathbf{u}^T/3  \right) I \\
I&=&\left(\int_{0}^{\infty} dr' \int_{-\infty}^0 ds  -2\int_{0}^{\infty} dr' \int_0^\infty ds \right) f(r,s)+\cc
\end{eqnarray}
Now we will add and subtract the quantity $\int_{0}^{\infty} dr \int_0^{\infty} ds \ f(r,s)+\cc $ to $I$ to get
\begin{eqnarray}
I=\left(\int_{0}^{\infty} dr \int_{-\infty}^0 ds \ f(r,s) +\int_{0}^{\infty} dr \int_{0}^{\infty} ds \ f(r,s) +\cc \right)- \left( 3\int_{0}^{\infty} dr \int_{0}^{\infty} ds \ f(r,s)+\cc\right)=I_1-I_2.
\end{eqnarray}
Notice that the term in parenthesis, $I_1$, can be evaluated since:
\begin{eqnarray}
I_1&=&\int_{0}^{\infty} dr \int_{-\infty}^0 ds \ f(r,s) +\int_{0}^{\infty} dr \int_{0}^{\infty} ds \ f(r,s) +\cc=\int_{-\infty}^{\infty} dr \int_{-\infty}^{\infty} ds \ f(r,s)\\
&=&\frac{4 \pi s_a s_b \tau^2}{3 R^2} \exp\left(-\frac{4 \tau^2 \eta_{ab}^2}{3 R^4} \mathbf{u} \mathbf{W} \mathbf{u}^T \right), \quad 
\mathbf{W}=\left(
\begin{array}{cc}
 2 \mu _a^2 & -\mu ^2 \\
 -\mu ^2 & 2 \mu _b^2
\end{array}
\right).
\end{eqnarray}
As for $I_2$ we can proceed as follows, 
\begin{eqnarray}
I_2=3\int_{0}^{\infty} dr \int_{0}^{\infty} ds \ f(r,s)+\cc=6 \int_{0}^{\infty} dr \int_{0}^{\infty} ds \exp\left(\mathbf{x}^T \mathbf{A} \mathbf{x} \right) \cos \left( \mathbf{v}^T \mathbf{x} \right).
\end{eqnarray}
But now we can let $s \to 2  s_a \tau q/\sqrt{3}$ and $r \to 2 s_b \tau p/\sqrt{3}$ to get:
\begin{eqnarray}
I_2=8 s_a s_b \tau^2 \int_0^{\infty} dq \int_0^{\infty} dp \exp\left(-(p,q)\mathbf{M}(p,q)^T \right) \cos\left(4 \tau \eta_{ab} (\delta \omega_a q+\delta \omega_b p)/\sqrt{3} \right).
\end{eqnarray}
Now we can put all these results together to obtain
\begin{eqnarray}
J_3= \frac{ \sqrt{\pi}  \tau  \varepsilon^3 }{2  s_a s_b} \exp \left(- \mathbf{u} \mathbf{N}\mathbf{u}/3  \right)&& \bigg\{ \frac{4 \pi s_a s_b \tau^2}{3 R^2} \exp\left(-\frac{4 \tau^2 \eta_{ab}^2}{3 R^4} \mathbf{u} \mathbf{W} \mathbf{u}^T \right)\\
&&-8 s_a s_b \tau^2 \int_0^{\infty} dq \int_0^{\infty} dp \exp\left(-(p,q)\mathbf{M}(p,q)^T \right) \cos\left(\frac{4 \tau \eta_{ab} (\delta \omega_a q+\delta \omega_b p)}{\sqrt{3}} \right) \bigg\} \nonumber.
\end{eqnarray}
Simplifying and noting that $\mathbf{N}/3+\mathbf{W}=\mathbf{Q}/R^4$ we obtain
\begin{eqnarray}
J_3&=& \frac{2\pi^{3/2}  \tau^3\varepsilon^3}{3R^2} \exp\left(-\mathbf{u}\mathbf{Q}\mathbf{u}^T/R^4 \right)\\
&&-4 \sqrt{\pi} \tau^3 \varepsilon^3  \exp \left(- \mathbf{u} \mathbf{N}\mathbf{u}^T/3  \right) \int_0^{\infty} dq \int_0^{\infty} dp \exp\left(-(p,q)\mathbf{M}(p,q)^T \right) \cos\left(4 \tau \eta_{ab} (\delta \omega_a q+\delta \omega_b p)/\sqrt{3} \right)\nonumber.
\end{eqnarray}
This last equation can be rewritten as Eqs. (8-11) using the definitions of Eq. (12) in the main text.
We now obtain a bound for $J_3$. First note that $|J_{3}|\leq |W|+|VZ|=W+|V|Z$; the only term that is hard to bound from the last inequality is $|V|$, and to this end we note the following chain of inequalities:
\begin{eqnarray}
\exp\left( -(p,q)\mathbf{M}(p,q)^{T}\right) \left|\cos \left( 4\tau \eta _{ab}(\delta \omega _{a}q+\delta \omega _{b}p)/\sqrt{3}\right)\right| &\leq& \exp
\left( -(p,q)\mathbf{M}(p,q)^{T}\right) \\
\int_{0}^{\infty }dp\int_{0}^{\infty }dq \exp\left( -(p,q)\mathbf{M}(p,q)^{T}\right) \left|\cos \left( 4\tau \eta _{ab}(\delta \omega _{a}q+\delta \omega _{b}p)/\sqrt{3}\right)\right| &\leq& \int_{0}^{\infty }dp\int_{0}^{\infty }dq  \exp\left( -(p,q)\mathbf{M}(p,q)^{T}\right).
\end{eqnarray}
Now note that for any function $g(p,q)$, $|\int_{0}^{\infty }dp\int_{0}^{\infty }dq \  g(p,q)|\leq \int_{0}^{\infty }dp\int_{0}^{\infty }dq |g(p,q)|$ and thus we can finally write
\begin{eqnarray*}
|V(\omega_a,\omega_b)|&=&\left| \int_{0}^{\infty }dp\int_{0}^{\infty }dq \exp\left( -(p,q)\mathbf{M}(p,q)^{T}\right) \cos \left( 4\tau \eta _{ab}(\delta \omega _{a}q+\delta \omega _{b}p)/\sqrt{3}\right)\right| \nonumber \\
&\leq &\int_{0}^{\infty }dp\int_{0}^{\infty }dq  \exp\left( -(p,q)\mathbf{M}(p,q)^{T}\right)=V(\bar \omega_a,\bar \omega_b).
\end{eqnarray*}
The right hand side of the last inequality can be easily evaluated
\begin{eqnarray}
V(\bar \omega_a,\bar \omega_b)=\begin{cases}
\frac{\pi +\tan ^{-1}(R^{2}/\mu ^{2})}{2R^{2}},\text{ if }\mu ^{2}<0 \\ 
\frac{\tan ^{-1}(R^{2}/\mu ^{2})}{2R^{2}},\text{ if }\mu ^{2}>0.
\end{cases}
\end{eqnarray}
Note from Eq. (12) of the main text that $R^4 \geq 3\mu^4$ this immediately implies that $V(\bar{\omega}_{a},\bar{\omega}_{b})<\pi /(3R^{2})$ and the general bound
\begin{eqnarray}
|J_3(\bar{\omega}_{a},\bar{\omega}_{b})| < \frac{2\pi ^{3/2}\varepsilon
^{3}\tau ^{3}}{3R^{2}} \left( \exp \left( -\mathbf{u}\mathbf{Q}\mathbf{u}^{T}/R^{4}\right)+2\exp \left( -\mathbf{u}\mathbf{N}\mathbf{u}^{T}/3\right) \right) \leq \frac{2\pi ^{3/2}\varepsilon^{3}\tau ^{3}}{R^{2}},
\end{eqnarray}
which is Eq. (15)
The last bound is obtained by taking $\mathbf{u}=\mathbf{0}$. Note that $\max_{\eta_a,\eta_b} \tau^2/R^2=1/\sqrt{3}$ is achieved for $\eta_a=\eta_b=0$. Nevertheless note that for this parameter values ($\eta_{ab}=\eta_a-\eta_b=0$) $J_3$ is identically zero. To show this note that if $\eta_{ab}=0$ then $R^4=3 \mu^4$, $M^4=\mu^4$, $\mu^2>0$, $\mathbf{Q}/R^4=\mathbf{N}/3$ and
\begin{eqnarray}
V(\omega_a,\omega_b)=V(\bar \omega_a,\bar \omega_b)=\frac{\pi}{6 R^2}.
\end{eqnarray}

\end{widetext}

\bibliographystyle{plain}
\bibliography{shortv2}

\end{document}